\documentclass{article}

\def\n#1{\left| #1 \right|}
\def\nn#1{\| #1 \|}
\newtheorem{theorem}{Theorem}[section]
\newtheorem{prop}{Proposition}[section]

\def\open#1{\setbox0=\hbox{$#1$}
\baselineskip = 0pt
\vbox{\hbox{\hspace*{0.4 \wd0}\tiny $\circ$}\hbox{$#1$}} 
\baselineskip = 10pt\!}

\def\opens#1{\setbox1=\hbox{${\scriptstyle #1}$}
\baselineskip = 0pt
\vbox{\hbox{\hspace*{0.4 \wd1} $\kern-0.35em {\scriptscriptstyle \circ}$}
\hbox{${\scriptstyle #1}$}} 
\baselineskip = 10pt\!}
 
\def\fn{\open{f}\,}

\def\li{\open{\lambda}\,}
\def\mi{\open{\mu}\,}
\def\lis{\opens{\lambda}}
\def\mis{\opens{\mu}}
\def\ri{\open{\rho}\,}

\def\dt{\partial_t}
\def\dtheta{\partial_\theta}
\def\dw{ \partial_w}
\def\supp{\mbox{\rm supp}} 
\def\R{{\rm I\kern-.1567em R}}

\begin{document} 
\title{On future geodesic completeness for 
       the Einstein-Vlasov system with hyperbolic symmetry}
\date{}
\author{Gerhard Rein\\
        Institut f\"ur Mathematik, Universit\"at Wien\\
        Strudlhofgasse 4, A-1090 Vienna, Austria
        }

\maketitle

\begin{abstract}
Spacetimes with collisionless matter evolving
from data on a compact Cauchy surface with hyperbolic symmetry
are shown to be timelike and null geodesically complete in the expanding
direction, provided the data satisfy a certain size restriction.
      
\end{abstract}

\section{Introduction}

Consider the Einstein-Vlasov system with hyperbolic symmetry.
For background information on this system which describes
the evolution of a collisionless gas in general relativity we refer
to \cite{RR}, for the notion of hyperbolic symmetry we refer to
\cite{ARR}.
As shown in \cite{ARR} the corresponding spacetime manifold can
be covered by areal coordinates in which the metric takes the form
\begin{equation} \label{metric}
ds^2 = - e^{2\mu} dt^2 + 
e^{2\lambda} d\theta^2 + t^2( d\psi^2 + \sinh^2 \psi\, d\phi^2),
\end{equation}
where $\mu$ and $\lambda$ are functions of $t$ and $\theta$,
periodic in $\theta$ with period $1$, and the areal time coordinate $t$ 
takes all values in $]R_0,\infty[$ for some $R_0\geq 0$. 
The main motivation of the latter investigation was to provide a 
set-up in which the global properties of solutions of 
the Einstein equations with hyperbolic symmetry coupled
to a matter field, in this case a collisionless gas, can be studied.
In the present note we give sufficient conditions on initial data
such that the spacetime manifold is 
timelike and null geodesically complete towards the future, i.~e.,
in the expanding direction. In the contracting direction such a spacetime
has a spacetime singularity, cf.\ 
\cite{ARR} or \cite{R}. In the spherically symmetric, 
asymptotically flat case of the Einstein-Vlasov system
geodesic completeness  was obtained in \cite{RR} for small data.
No analogous result exists for cosmological models.

The geodesic equations for a metric of the 
form (\ref{metric})
imply that along geodesics the variables $t,\ \theta,
\ p^0$, and
\[
w := e^\lambda p^1,\ L:= t^4 ((p^2)^2 + \sinh^2 \psi\, (p^3)^2)
\]
satisfy the following system of differential equations:
\begin{equation} \label{cs1}
\frac{d\theta}{d\tau} =
e^{-\lambda} w,\ 
\frac{dw}{d\tau} = 
- \lambda_t p^0 w - e^{2\mu -\lambda} \mu_\theta (p^0)^2,\
 \frac{dL}{d\tau} = 0,
\end{equation}
\begin{equation} \label{cs2}
\frac{dt}{d\tau} = p^0,\
\frac{dp^0}{d\tau} =
- \mu_t (p^0)^2 - 2 e^{-\lambda} \mu_\theta p^0 w - e^{-2\mu}\lambda_t w^2 - 
e^{-2\mu} t^{-3} L.
\end{equation}
For a particle with rest mass $m$
moving forward in time $p^0$ can be expressed by the
remaining variables,
\[
p^0 = e^{-\mu} \sqrt{m^2+w^2 + L/t^2} > 0, 
\]
and the corresponding
geodesic as well as the solution of (\ref{cs1}), (\ref{cs2}) 
can be reparameterized by coordinate time $t$.
Consider an ensemble of such particles, all with rest mass equal
to unity. Due to the symmetry their density on the mass shell
\[
\{g^{\alpha \beta} p^\alpha p^\beta = -1,\ p^0 >0\}
\]
can be written as
\[
f=f(t,\theta,w,L),
\]
cf.\ \cite{ARR}. In these variables the Einstein-Vlasov system 
takes the form
\begin{equation} \label{av}
\dt f +\frac{ e^{\mu - \lambda} w}{\sqrt{1+w^2 + L/t^2}} \dtheta f -
\left( \lambda_t w + e^{\mu - \lambda} 
\mu_\theta \sqrt{1+w^2 + L/t^2} \right) \dw f =0,
\end{equation}
\begin{eqnarray} 
e^{-2\mu} (2 t \lambda_t + 1) - 1 
&=&
8\pi t^2 \rho , \label{af1}\\
e^{-2\mu} (2 t \mu_t - 1) + 1 
&=& 
8\pi t^2 p, \label{af2} 
\end{eqnarray}
\begin{equation}
\mu_\theta = 
- 4 \pi t e^{\mu + \lambda} j, \label{af3}
\end{equation}
\begin{equation}
e^{- 2 \lambda} \Bigl(\mu_{\theta \theta} + 
\mu_\theta (\mu_\theta - \lambda_\theta)\Bigr)
- e^{-2\mu}\Bigl(\lambda_{tt} + (\lambda_t + 1/t)(\lambda_t - \mu_t)\Bigr) 
=
4 \pi q, \label{af4}
\end{equation}
where
\begin{eqnarray}
\rho(t,\theta) 
&:=& 
\frac{\pi}{t^2} \int_{-\infty}^\infty \int_0^\infty
\sqrt{1+w^2 + L/t^2} f(t,\theta,w,L)\,dL\,dw,\label{ar}\\
p(t,\theta) 
&:=& 
\frac{\pi}{t^2} \int_{-\infty}^\infty \int_0^\infty
\frac{w^2}{\sqrt{1+w^2 + L/t^2}} f(t,\theta,w,L)\,dL\,dw, \label{ap}\\
j(t,\theta) 
&:=& 
\frac{\pi}{t^2} \int_{-\infty}^\infty \int_0^\infty
w f(t,\theta,w,L)\,dL\,dw, \label{aj}\\
q(t,\theta) 
&:=& 
\frac{\pi}{t^4} \int_{-\infty}^\infty \int_0^\infty
\frac{L}{\sqrt{1+w^2 + L/t^2}} f(t,\theta,w,L)\,dL\,dw.  \label{aq}
\end{eqnarray}
The characteristics of the Vlasov equation (\ref{av})
along which $f$ is constant are the solutions of 
\begin{equation} \label{cs}
\dot \theta =
\frac{e^{\mu-\lambda}w}{\sqrt{1+w^2+L/t^2}},\ 
\dot w = 
- \lambda_t w - e^{\mu-\lambda} \mu_\theta \sqrt{1+w^2+L/t^2},\ \dot L = 0
\end{equation}
where the dot denotes differentiation with respect to coordinate
time $t$ and the relation with (\ref{cs1}), (\ref{cs2}) is obvious.

Initial data are prescribed at
some time $t=t_0 >0$: 
\[
f(t_0,\theta,w,L) = \fn (\theta,w,L),\ \lambda (t_0,\theta) = \li (\theta),\ 
\mu (t_0,\theta) = \mi (\theta). 
\]
If these data are $C^1$ and 
periodic in $\theta$ with period $1$, if $\fn$ is compactly
supported with respect to $w$ and $L$, and if the data satisfy 
the constraint equation (\ref{af3}), then the corresponding solution exists
for all future areal time $t\geq t_0$, cf.\
\cite{ARR}. To save a little notation let $t_0=1$.

In the next section a bound on $w$ along characteristics of
(\ref{av}) is established
for restricted data. In Section~3 the lapse function
$e^\mu$ is estimated, again for restricted data. 
In the last section both estimates are combined to obtain future geodesic
completeness.
 
\section{An estimate along characteristics}
\setcounter{equation}{0}
Let
\begin{eqnarray*}
w_0 
&:=&
\sup \left\{ \n{w} \mid (r,w,L) \in \supp \fn \right \} < \infty,\\
L_0 
&:=&
\sup \left\{ L \mid (r,w,L) \in \supp \fn \right \} < \infty ;
\end{eqnarray*}
without loss of generality $w_0 >0,\ L_0 >0$.
For $t\geq 1$ define
\begin{eqnarray*}
P_+(t) 
&:=& 
\max \left\{0,\max \Bigl\{ w \mid (r,w,L) \in \supp f(t) \Bigr\}\right\},\\
P_-(t) 
&:=& 
\min \left\{0,\min \Bigl\{ w \mid (r,w,L) \in \supp f(t) \Bigr\}\right\}.
\end{eqnarray*}
We claim that if
\begin{equation} \label{small1}
4 \pi^2 (1+L_0) L_0 \nn{\fn}_\infty < 1
\end{equation}
then
\begin{equation} \label{pbound}
P_+(t) \leq w_0 \sqrt{t},\ P_-(t) \geq -w_0 \sqrt{t},\ t \geq 1.
\end{equation}
Assume the estimate on $P_+$ were false for some $t$.
Define
\[
t_0:=\sup\{t\geq 1 | P_+(s) \leq w_0 \sqrt{s},\ 1 \leq s \leq t\}
\]
so that $1\leq t_0 < \infty$ and $P_+(t_0)=w_0 \sqrt{t_0} >0$. 
Choose $\epsilon \in ]0,1[$ such that
\[
4 \pi^2 (1+L_0) L_0 \nn{\fn}_\infty \leq (1-\epsilon)^2.
\]
By continuity, there exists some $t_1>t_0$ such that the following holds:
\[
(1-\epsilon) P_+(s) > 0,\ s \in [t_0,t_1],
\]
and if for some characteristic curve $(\theta(s),w(s),L)$ in the support
of $f$, that is, with $(\theta(1),w(1),L)\in \supp \fn$,
and for some $t \in ]t_0,t_1]$ the estimate
\begin{equation} \label{charlarge}
(1-\epsilon/2) P_+(t)\leq w(t) \leq P_+(t)
\end{equation}
holds, then 
\begin{equation} \label{charstillarge}
(1-\epsilon) P_+(s)\leq w(s) \leq P_+(s),\ s \in [t_0,t];
\end{equation}
note that the estimates on $w$
from above hold by definition of $P_+$ in any case.

Let $(\theta(s),w(s),L)$ be a characteristic in the support of $f$
satisfying (\ref{charlarge}) for some $t \in ]t_0,t_1]$ and thus 
(\ref{charstillarge}) on $[t_0,t]$.
Then on $[t_0,t]$,
\begin{eqnarray*}
\dot w
&=&
\frac{4 \pi^2}{s} e^{2 \mu} \int_{-\infty}^\infty \int_0^\infty
\left( \tilde w  \sqrt{1 + w^2 + L /s^2} 
-w \sqrt{1 + \tilde w^2 + \tilde L /s^2} \right)
f \, d\tilde L\, d\tilde w \\
&&
{}+ \frac{1 - e^{2 \mu}}{2 s} w \\
&\leq&
\frac{4 \pi^2}{s} e^{2 \mu} \int_0^{P_+(s)}\int_0^{L_0}
\frac{\tilde w^2  (1 + w^2 + L /s^2) 
- w^2 (1 + \tilde w^2 + \tilde L /s^2)}
{\tilde w  \sqrt{1 + w^2 + L /s^2} 
+ w \sqrt{1 + \tilde w^2 + \tilde L /s^2}}
f \, d\tilde L\, d\tilde w \\
&&
{}+ \frac{1 - e^{2 \mu}}{2 s} w \\
&\leq&
\frac{4 \pi^2}{s} e^{2 \mu} \int_0^{P_+(s)}\int_0^{L_0}
\frac{\tilde w (1+L)}{w}
f \, d\tilde L\, d\tilde w 
+ \frac{1 - e^{2 \mu}}{2 s} w \\
&\leq&
4 \pi^2 L_0 (1+L_0) \nn{\fn}_\infty \frac{e^{2\mu}}{2 s}P_+^2(s) \frac{1}{w}
+ \frac{1 - e^{2 \mu}}{2 s} w\\
&\leq&
\frac{e^{2\mu}}{2 s}\left((1-\epsilon)P_+(s)\right)^2 \frac{1}{w} 
+ \frac{1 - e^{2 \mu}}{2 s} w \leq
\frac{e^{2\mu}}{2 s} w + \frac{1 - e^{2 \mu}}{2 s} w =
\frac{1}{2s} w.
\end{eqnarray*}
Thus
\[
w(t) \leq w(t_0) \sqrt{t/t_0} \leq P_+(t_0)\sqrt{t/t_0} = w_0 \sqrt{t}
\]
by assumption on $t_0$. This estimate holds only for 
characteristics which
satisfy (\ref{charlarge}), but this is sufficient to conclude that
\[
P_+(t)\leq w_0 \sqrt{t},\ t \in [t_0,t_1],
\]
in contradiction to the choice of $t_0$. The estimate on $P_+$ is now
established. 

The analogous arguments for characteristics with $w<0$
yield the assertion for $P_-$, and we have shown:

\begin{prop} \label{wbound}
For any solution of the Einstein-Vlasov system with hyperbolic symmetry
written in areal coordinates and with initial data as above, 
\[
|w| \leq w_0 \sqrt{t},\ (\theta,w,L) \in \supp f(t),\ t \geq 1,
\]
provided the data satisfy the size restriction
\[
4 \pi^2 (1+L_0) L_0 \nn{\fn}_\infty < 1.
\]
\end{prop}

{\bf Remark}: The estimate above is sufficient to 
show that all geodesics which correspond to characteristics in the support
of $f$ exist for all proper time in the expanding direction, but 
geodesic completeness requires the same to hold 
for timelike geodesics in general.

\section{An estimate for the lapse function}
\setcounter{equation}{0}

We reconsider some of the estimates in \cite{ARR}
from the point of view of getting a better
bound on $e^{\mu}$ for restricted data. 
First recall that by integration of the 
field equation (\ref{af2}) and since $p$
is non-negative,
\begin{equation}\label{loweste2mu}
e^{2 \mu(t,\theta)} 
\geq
\frac{t}{e^{-2\mis(\theta)}-1+t}
\geq 
\frac{t}{c_0-1+t},\ t \geq 1,
\end{equation}
where
\[
c_0:= \max_{\theta \in [0,1]}e^{-2\mi(\theta)}.
\]
Next we recall that by a lengthy computation,
\begin{eqnarray*}
\frac{d}{dt}
\int_0^1 e^{\mu + \lambda} \rho(t,\theta)\, d\theta
&=&
- \frac{1}{t} \int_0^1 e^{\mu + \lambda}
\biggl[ 2 \rho + q - \frac{\rho + p}{2}(1 - e^{2 \mu})\biggr]\, 
d\theta \\
&\leq&
- \frac{1}{t} \int_0^1 e^{\mu + \lambda}
\biggl[ 2 \rho  - \frac{\rho + p}{2}\biggr]\, d\theta
- \frac{1}{c_0-1+t} \int_0^1 e^{\mu + \lambda} 
\frac{\rho + p}{2}\, d\theta\\
&&
{}
\leq
- \frac{2}{c_1+t} \int_0^1 e^{\mu + \lambda} \rho \, d\theta
\end{eqnarray*}
where
\[
c_1:=\max\{0,c_0-1\}.
\]
Thus
\begin{equation} \label{decrhoint}
\int_0^1 e^{\mu + \lambda} \rho(t,\theta)\, d\theta \leq 
(c_1+1)^2 \int_0^1 e^{\mis + \lis} \ri(\theta)\, d\theta 
\; t^{-2},\ t \geq 1.
\end{equation}
By (\ref{loweste2mu}) and since $p \leq \rho$,
\[
\frac{\partial}{\partial t} e^{\mu - \lambda} =
e^{\mu - \lambda}
\left[ 4 \pi t e^{2 \mu} (p-\rho) + \frac{1- e^{2 \mu}}{t}\right]
\leq
e^{\mu - \lambda} \left(\frac{1}{t}-\frac{1}{e^{-2\mis}-1+t}\right)
\]
which implies that
\begin{equation} \label{emuminla}
e^{\mu(t,\theta) - \lambda(t,\theta)} \leq 
e^{\mis(\theta) - \lis(\theta)} 
\frac{e^{-2\mis(\theta)}t}{e^{-2\mis(\theta)}-1+t}
\leq e^{|\mis(\theta)| -\lis(\theta)},
\ t \geq 1,\ 
\theta \in [0,1].
\end{equation}
Now we estimate the average of $\mu(t)$, using (\ref{loweste2mu}),
(\ref{decrhoint}), (\ref{emuminla}):
\begin{eqnarray*}
\int_0^1 \mu(t,\theta)\, d\theta
&=&
\int_0^1 \mi(\theta)\, d\theta +
\int_{1}^t\int_0^1 \mu_t(s,\theta)\, d\theta\, ds\\
&=&
\int_0^1 \mi(\theta)\, d\theta + 4 \pi \int_{1}^t s
\int_0^1 e^{\mu -\lambda}e^{\mu +\lambda}p\, d\theta\, ds + 
\int_{1}^t \frac{1}{2s} \int_0^1 (1-e^{2\mu})\, d\theta\, ds \\
&\leq&
C +
4 \pi \max_{\theta \in [0,1]} e^{|\mis(\theta)| -\lis(\theta)} (c_1+1)^2
\int_0^1 e^{\mis +\lis}\ri\, d\theta\ \ln t
\end{eqnarray*}
where the constant $C>0$ depends in a complicated but irrelevant 
way on the initial
data. If the data satisfy the size restriction 
\begin{equation} \label{small2}
4 \pi \max_{\theta \in [0,1]} e^{|\mis(\theta)| -\lis(\theta)} (c_1+1)^2
\int_0^1 e^{\mis +\lis}\ri\, d\theta \leq \frac{1}{2}
\end{equation}
then the estimate above becomes
\[
\int_0^1 \mu(t,\theta)\, d\theta \leq C + \frac{1}{2} \ln t,\ t \geq 1.
\]
Since
\[
\left|\mu (t,\theta) - \int_0^1 \mu(t,\sigma)\, d\sigma \right|
\leq C t^{-1},\ t \geq 1,\ \theta \in [0,1],
\]
cf.\ \cite[Eqn.~3.14]{ARR}, it follows that
\[
\mu(t,\theta) \leq C + \frac{1}{2} \ln t,\ t \geq 1.
\]
\begin{prop} \label{lapsebound}
For any solution of the Einstein-Vlasov system with hyperbolic symmetry
written in areal coordinates and with initial data as above,
\[
e^{\mu(t,\theta)} \leq C \sqrt{t},\ t \geq 1,\ \theta \in [0,1],
\]
with $C>0$ depending on the initial data, provided they 
satisfy the size restriction
\[
4 \pi \max_{\theta \in [0,1]} e^{|\mis(\theta)| -\lis(\theta)}\;
\max\{1,\max_{\theta \in [0,1]} e^{-4 \mis(\theta)}\}\;
\int_0^1 e^{\mis +\lis}\ri\, d\theta 
 \ \leq \frac{1}{2}.
\]
\end{prop}

\section{Geodesic completeness}
\setcounter{equation}{0}

Let $]\tau_-,\tau_+[ \ni \tau \mapsto (x^\alpha(\tau),p^\beta(\tau))$
be a geodesic whose existence interval is maximally extended
and such that $x^0(\tau_0)=t(\tau_0)=1$ for some $\tau_0 \in ]\tau_-,\tau_+[$.
We want to show that for timelike and null geodesics which
move forward in time, $\tau_+ = +\infty$.
Consider first the case of a timelike geodesic, i.~e.,
\[
g_{\alpha \beta} p^\alpha p^\beta = -m^2,\ p^0 >0
\]
with $m>0$. Since $dt/d\tau = p^0 >0$
the geodesic can be parameterized by the coordinate time $t$.
With respect to coordinate time the geodesic exists on the interval
$[1,\infty[$ since on bounded $t$-intervals the Christoffel symbols
are bounded and the right hand sides of the geodesic equations written
in coordinate time are linearly bounded in $p^1, p^2, p^3$. Along the geodesic
we define $w$ and $L$ as above. Then the relation
between coordinate time and proper time along the geodesic is given by
\[
\frac{dt}{d\tau} = p^0 = e^{-\mu}\sqrt{m^2 + w^2 + L/t^2},
\]
and to control this we 
need to control $w$ as a function of coordinate time. 
Consider first the case where 
$w(t)>0$ for some $t>1$ and define $t_0 \geq 1$ minimal with the property that
$w(s) >0$ for $s\in ]t_0,t]$. We argue similarly to
Section~2, making use of the additional estimates which
we have established under the size restrictions on the initial data,
cf.\ Props.~(\ref{wbound}) and (\ref{lapsebound}):
\begin{eqnarray*}
\dot w
&=&
\frac{4 \pi^2}{s} e^{2 \mu} \int_{-\infty}^\infty \int_0^\infty
\left( \tilde w  \sqrt{m^2 + w^2 + L /s^2} 
-w \sqrt{1 + \tilde w^2 + \tilde L /s^2} \right)
f \, d\tilde L\, d\tilde w \\
&&
{}+ \frac{1 - e^{2 \mu}}{2 s} w \\
&\leq&
C\; \int_0^{P_+(s)}\int_0^{L_0}
\frac{\tilde w^2  (m^2 + w^2 + L /s^2) 
- w^2 (1 + \tilde w^2 + \tilde L /s^2)}
{\tilde w  \sqrt{m^2 + w^2 + L /s^2} 
+ w \sqrt{1 + \tilde w^2 + \tilde L /s^2}}
f \, d\tilde L\, d\tilde w + \frac{w}{2 s} \\
&\leq&
C\; \int_0^{P_+(s)}\int_0^{L_0}
\frac{\tilde w (m^2+L)}{w}
f \, d\tilde L\, d\tilde w 
+ \frac{w}{2 s} \leq
\frac{C s}{w}
+ \frac{w}{2 s}
\end{eqnarray*}
or
\[
\frac{d}{ds} w^2 \leq C s + \frac{1}{s} w^2 .
\]
This implies that
\[
w^2(t) \leq \frac{t}{t_0} w^2(t_0) + C t (t-t_0).
\]
Now either $t_0>0$ in which case $w(t_0)=0$ or $t_0=1$.
In both cases $w(t) \leq C t$ where the positive constant $C$
depends on the initial data of the solution of the Einstein-Vlasov system
and on the initial data of the particular geodesic. An analogous
argument applies if $w(t) < 0$ for some $t> 1$. Thus along the geodesic,
\[
\frac{d\tau}{dt} = \frac{e^\mu}{\sqrt{m^2 + w^2 + L/t^2}}
\geq \frac{C}{\sqrt{m^2 + C t^2 + L}}.
\]
Since the integral of the right hand side over $[1,\infty[$ diverges,
$\tau_+ = + \infty$ as desired.
 
Now consider a null geodesic which moves forward in time initially, 
i.~e., $m=0$ and $p^0(\tau_0)>0$. The quantity $L$ is again
conserved. In particular, if $L>0$ then
$p^0$ remains positive on the maximal existence interval of the geodesic.
If $L=0$ the same is true since otherwise
$p^0(\tau)=0$ and necessarily also $p^1(\tau)=0$ at some $\tau$ which
by uniqueness of the solutions of the geodesic
equations implies that $p^0=p^1=0$ always, a contradiction.
The argument can now be carried out exactly as before,
implying that $\tau_+ = + \infty$.

\begin{theorem} 
If the initial data of a solution of the Einstein-Vlasov system
with hyperbolic symmetry, written
in areal coordinates, satisfy the size restrictions stated in
Props.\ (\ref{wbound}) and (\ref{lapsebound}), then the 
corresponding spacetime is timelike and null geodesically complete in the
expanding direction.
\end{theorem}

{\bf Final remarks.} In \cite{CB}, conditions on the spacetime
metric, the gradient of the lapse, and the extrinsic curvature
are given which imply that the spacetime is timelike and null 
geodesically complete. These conditions
seem stronger than what we needed to establish geodesic completeness
in the present note. Among other things, in \cite{CB}
the lapse function needs to be uniformly 
bounded from above and away from zero which is more than what our estimates
provide. 

The control on the $w$ component of the support of $f$
obtained in Section 2 implies the following estimates
for the source terms in the field equations:
\[
\rho(t,\theta),\ p(t,\theta),\ |j(t,\theta)|
\leq C t^{-1},\ q(t,\theta)\leq C t^{-7/2},\ t \geq 1,\ \theta \in [0,1].
\]
If $R_{\alpha\beta\gamma}^{\phantom{\alpha\beta\gamma}\delta}$
denotes the Riemann curvature tensor then the quantity
\[
K(t,\theta):=
\left(R_{\alpha\beta\gamma\delta} R^{\alpha\beta\gamma\delta}\right) 
(t,\theta)
\]
is known as the Kretschmann scalar.
In \cite{R} it was shown that
\begin{eqnarray*}
K 
&=&
4 \left( e^{- 2 \lambda} \left(\mu_{\theta\theta} + 
\mu_\theta (\mu_\theta - \lambda_\theta)\right)
- e^{-2\mu}\left(\lambda_{tt} + \lambda_t (\lambda_t - \mu_t)\right) 
\right)^2 \\
&&
+ \frac{8}{t^2} \left( e^{-4\mu} \lambda_t^2 + e^{-4 \mu} \mu_t^2
- 2 e^{-2(\lambda + \mu)} \mu_\theta^2 \right) 
+ \frac{4}{t^4} \left( e^{-2 \mu} -1 \right)^2
\end{eqnarray*}
and a lower bound for this quantity was established
which is positive and blows up as $t \to 0$. Using the
bounds established above one can show that
$K$ decays like $t^{-2}$ for $t \to \infty$.

The question whether the spacetime is
geodesically complete for data without a size restriction
remains open.

{\em Acknowledgement:} The auther acknowledges support by the Wittgenstein 
2000 Award of P.~A.~Markowich.

\end{document}